# Quantum-dot single-electron transistor as thermoelectric quantum detectors at terahertz frequencies

Mahdi Asgari[1], Dominique Coquillat[2], Guido Menichetti[3,4] Valentina Zannier[1], Nina Dyakonova[2], Wojciech Knap[2,5], Lucia Sorba[1], Leonardo Viti[1*], Miriam Serena Vitiello[1*].

[1]NEST, CNR—Istituto Nanoscienze and Scuola Normale Superiore, Piazza San Silvestro 12, 56127, Pisa, Italy
[2]Laboratoire Charles Coulomb UMR 5221 CNRS-Université Montpellier, Place Eugène Bataillon CC074, F-34095, Montpellier, France
[3] Istituto Italiano di Tecnologia, Graphene Labs, Via Morego 30, I-16163 Genova, Italy
[4]Dipartimento di Fisica dell'Universit di Pisa, Largo Bruno Pontecorvo 3, I-56127 Pisa, Italy
[5]CENTERA Laboratories, Institute of High Pressure Physics, Polish Academy of Sciences, 01-142 Warsaw, Poland

*miriam.vitiello@sns.it
*leonardo.viti@nano.cnr.it

**Low dimensional nano-systems are promising candidates for manipulating, controlling and capturing photons with large sensitivities and low-noise. If quantum engineered to tailor the energy of the localized electrons across the desired frequency range, they can allow devising efficient quantum sensors across any frequency domain. Here, we exploit the rich few-electrons physics to develop millimeter-wave nanodetectors employing as sensing element an InAs/InAs$_{0.3}$P$_{0.7}$ quantum-dot nanowire, embedded in a single electron transistor. Once irradiated with light the deeply localized quantum element exhibits an extra electromotive force driven by the photothermoelectric effect, which is exploited to efficiently sense radiation at 0.6 THz with a noise equivalent power < 8 pWHz$^{-1/2}$ and almost zero dark current. The achieved results open intriguing perspectives for quantum key distributions, quantum communications and quantum cryptography at terahertz frequencies.**

Applications in quantum information technology usually require a tight control of the orbitals states of a single electron, needed to encode and transfer information with high fidelity[1]. This has recently stimulated a wide interest in efficient, low-noise receivers that can detect a controlled photons number (photon counters), or even single-photons[2] - the ultimate limit of detection sensitivity - in quantum nanostructures. This has long been inaccessible for terahertz (THz) or gigahertz waves (wavelengths in the range 60 μm-1mm) because of the small energy quanta, as small as a part per one thousand of the photon energies in the visible or near-infrared regions. However, the far-infrared region of the electromagnetic spectrum discloses a peculiar potential in this respect: it is a rich domain of spectroscopy and metrological research, and a frontier region for a variety of applications in biomedical diagnostics[3], quality and process controls[4], security[5], high data rate wireless communication[6] as well as for applications in optical quantum cryptography[7], and for quantum key distribution[8].

Quantum detectors, conventionally employed in the visible, near-infrared or mid-infrared regimes - including photodiodes[9], photoconductors[10], phototransistors[11], charge-coupled detectors (CCDs)[12], photomultiplier tubes[13] or semiconductor quantum-well infrared photodetectors (QWIPs)[14] - are devices that convert incoming photons directly into an electrical signal, as opposed to thermal detectors that rely on the conversion of incoming radiation to heat[15]. QWIPs represent the benchmark technology for quantum applications in the mid-infrared frequency range[16], owing to the high sensitivity and ultrafast response times, and the short lifetime of the intersubband transitions ($\tau_{IST} \sim$ ps)[15]. Moreover, intersubband transitions in quantum wells (QWs) go along with pronounced optical nonlinearities, resulting in huge nonlinear coefficients for second harmonic generation, more than three orders of magnitude larger than for the host material GaAs[17]. Pushing the operation of QWIPs in the far-infrared is however extremely challenging due to the low energy of far-infrared photons. Only few reports at high THz frequencies (4.5-7 THz) are presently available either in a standard mesa configuration, normally substrate-coupled through a polished facet[16], or in a double metal patch-antenna array architecture[18], or in a metamaterial configuration with dimensions below the diffraction limit[19], at 3 THz, and with a maximum speed of 3 GHz in an array configuration[20].

Quantum dots infrared photodetectors (QDIPs) represent a promising route to overcome some of the major limitations of THz QWIPs[21]. In particular, the energy level configuration and the orbital occupation can be controlled via the QD diameter-height and the gate bias, respectively. QDs are also inherently sensitive to normal incidence photoexcitation, therefore do not require 45° polished facets, metal-gratings[22], engineered band mixing[22,23] or reflectors, as it is conventionally needed for QWIPs[23]. Owing to the discrete density of states, they are well suited for tunable narrow band detection. Most importantly, the three-dimensional confinement leads to phonon bottleneck effect[24], inhibiting phonon scattering in QDs as compared to QWs[23,25] and increasing the lifetime of photoexcited carriers (~ 5-10 ps)[23], whose phonon-mediated relaxation to the ground state is eventually hindered in favor of an enhanced probability of tunneling out of the dot. This effect is then expected to give rise to a more efficient detection, as a consequence of a larger quantum efficiency and to enable operation at higher temperatures (up to 50-60 K)[21,23]. The reduced dependence of the density of states on temperature and the longer carrier lifetime (one to two order of magnitude) in QD have the additional advantage of reducing the dark current with respect to QWIPs[26].

QDIPs are promising candidates for applications in THz communication[15,27,28]. Once implemented in a single electron transistor (SET) geometry, a low noise equivalent power (NEP) $\sim 10^{-19}$ WHz$^{-1/2}$ [15,28] can be indeed engineered under a precise control of bias[15]. When illuminated with a radiation energy that is not in resonance with the QD intersubband transition, QD-based devices still exhibit good detection performances (e.g. responsivity up to 100 A/W[29]), owing to the inherently strong non-linearity of the current-voltage characteristics.

Here, we conceive and devise quantum-dot millimeter-wave nanodetectors employing InAs/InAsP QD nanowires (NWs) that, thanks to the small effective mass and favorable Fermi level pinning[30,31], give rise to localized QDs characterized by large charging energy[31,32]. Employing the confinement defined by the double-barrier heterostructure, we engineer a QD SET that, once irradiated with light, exhibits an extra electromotive force driven by the photothermoelectric (PTE) effect, which can be exploited to efficiently sense the incoming radiation with NEP levels down to 8 pWHz$^{-1/2}$. Importantly, the demonstrated PTE quantum detectors operate under zero-bias, therefore the dark current is largely reduced with respect to standard configurations employing biased systems[16,18,19].

In the last years, heterostructured semiconductor NWs proved to be a promising technological platform[33] for devising sensitive, high-speed, low-noise detectors across the THz [34]. NW field effect transistors (FETs) with controlled and tailored composition[26], are compatible with on-chip technologies[33] and are featured with a characteristic attofarad-order capacitance that makes them well suited for low capacitance integrated circuits[35,36]. Even though the stoichiometric and geometric control in the growth of axially heterostructured NWs can allow tailoring tunnel barrier properties on purpose, it does not allow a widely tunable tunnel coupling, as compared to electrostatically defined structures[37]. This provide an important benefit for efficient thermoelectric conversion[31,38], or for single-photon QD detectors which may require a broad different range of tunneling rates [39]. An alternative, more efficient strategy to simultaneously optimize charge stability and tunneling relies on engineering and tuning electrostatically the orbital configuration within the QD[40].

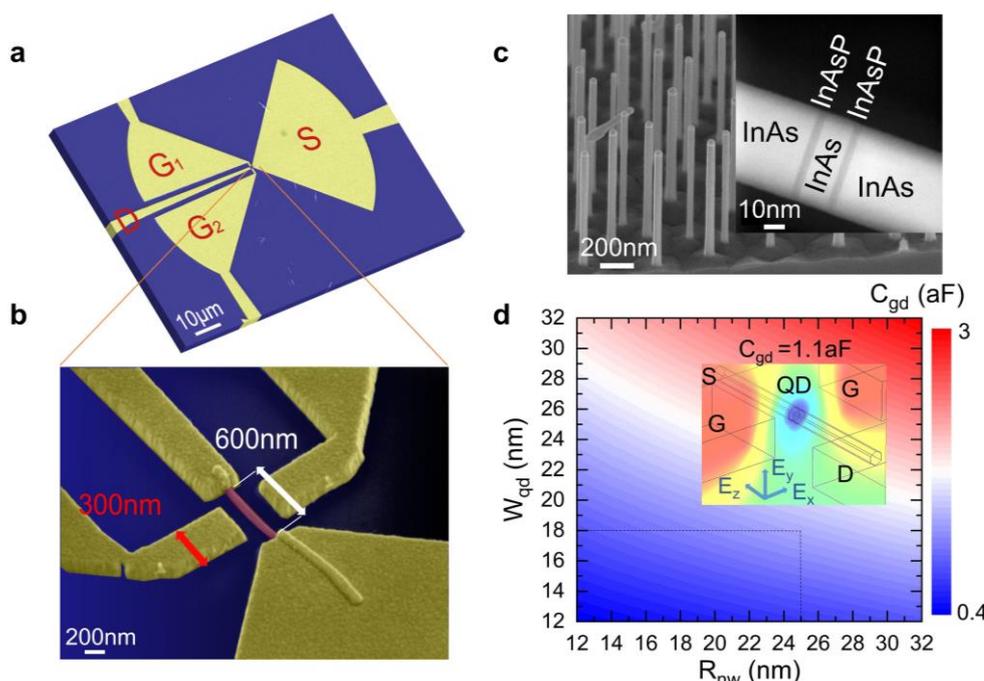

**Figure 1 a)** Scanning electron microscopy (SEM) image of a planar on-chip split bow-tie antenna. One side of the antenna is connected to the source electrode while the opposite side is connected to the arms of double lateral gate contacts. **b)** SEM image of a prototypical quantum dot nanowire (QD-NW) single electron transistor (SET). **c)** SEM image of a forest of epitaxially grown InAs nanowires, (inset) Scanning transmission electron microscopy (STEM) image of a single InAs/InAsP QD-NW. **d)** Color map of the gate-QD capacitance ($C_{gd}$) as a function of QD axial dimension (distance between the barriers, $W_{qd}$) and NW radius ($R_{nw}$), calculated numerically using an electrostatic simulation of the QD-NWFET performed with a commercial software (COMSOL Multiphysics). Inset: three-dimensional image of the SET channel overlaid to the simulated distribution of electrostatic potential around the InAs QD. The simulated $C_{gd}$ value is 1.1aF for our specific device geometry ($W_{qd}$ = 18nm, $R_{nw}$ = 25nm).

We here exploit InAs/InAs$_{0.3}$P$_{0.7}$ QD-NWs grown via a gold assisted chemical-beam-epitaxy (CBE). This material system allows combining semiconductors with different lattice parameters in axial heterostructures, thanks to the efficient strain relaxation along the NW sidewalls. Moreover, the InAs/InP system is particularly suitable for the realization of high-quality axial NW heterostructures like QDs and superlattices in Au-assisted growth. Indeed, the very low solubility of both As and P into Au allows to obtain atomically sharp interfaces in both growth directions[41]. However, in the case of NWs grown from a metal seed nanoparticle (NP) by the vapor-liquid-solid (VLS) mechanism, the chemical composition of the NP changes when the growth is switched from one material to the other and this strongly affects the NP stability, the growth mode (straight or kinked) and the growth rate[41,42]. In particular, the growth of alternating InAs/InP segments is prone to nucleation delay during the growth of the InP segment, that can lead to NP reconfiguration, which, in turn, affects the growth dynamics[42]. On the other hand, if InAs$_{(1-x)}$P$_x$ alloys instead of InP are grown on top of InAs NWs, the nucleation delay is not present. As a consequence, the growth of InAs/InAs$_{(1-x)}$P$_x$ heterostructures is uniform and very symmetric thicknesses are obtained for the same growth times. Finally, the height of the tunneling barriers can be tuned by changing the P/As ratio in the alloy segments[43].

A 18 nm long segment of InAs with band gap $E_g = 0.40$ eV and electron effective mass $m^* = 0.063 m_e$[44], where $m_e$ is the free electron mass, is confined by thin (5 ± 2 nm) $InAs_{0.3}P_{0.7}$ barriers with relatively high $E_g = 1.03$ eV and $m^* = 0.067 m_e$[45], leading to quantum confinement along the NW axial direction. The resulting $InAs/InAs_{0.3}P_{0.7}$ QD-NWs are transferred from the growth substrate over a 300 nm/350 μm $SiO_2$/intrinsic silicon wafer, where the detectors are nano-fabricated. Individual NWs are integrated in planar laterally-gated FETs[46] (Figs. 1a-1c), employing a combination of electron beam lithography (EBL) and thermal evaporation (see Supporting Information). Under this configuration the nano-system behaves like a few-electrons transistor and the electrical transport can be described within the framework of the constant interaction model[47]. The self-capacitance of the QD ($C_\Sigma$) defines the charging energy $\delta = e^2/C_\Sigma$ required for adding one electron to the dot. Together with the capacitance between the QD and the gate electrode ($C_{gd}$, Fig. 1d), it determines the gate lever arm, $\alpha_G = C_{gd}/C_\Sigma$, which is explicitly related to the capacitive coupling with the gate electrode. If $\delta \geq k_B T$, being $k_B$ the Boltzmann constant and T the temperature, the source (S) to drain (D) current is expected to exhibit a set of sharp peaks as a function of the gate voltage ($V_G$), corresponding to the resonant tunneling of single electrons through the QD, and reflecting Coulomb interactions between electrons[46]. In this Coulomb blockade regime[48], the voltage interval between consecutive peaks is defined by the sum of the energy levels spacing $\Delta E$ and the charging energy[48]. By properly choosing the geometrical parameters of the dot, i.e. NW radius ($R_{nw}$) and width of the InAs segment (Fig. 1c) between the two $InAs_{0.3}P_{0.7}$ barriers ($W_{qd}$), the distance between consecutive energy levels can be tailored to be resonant with a desired photon energy.

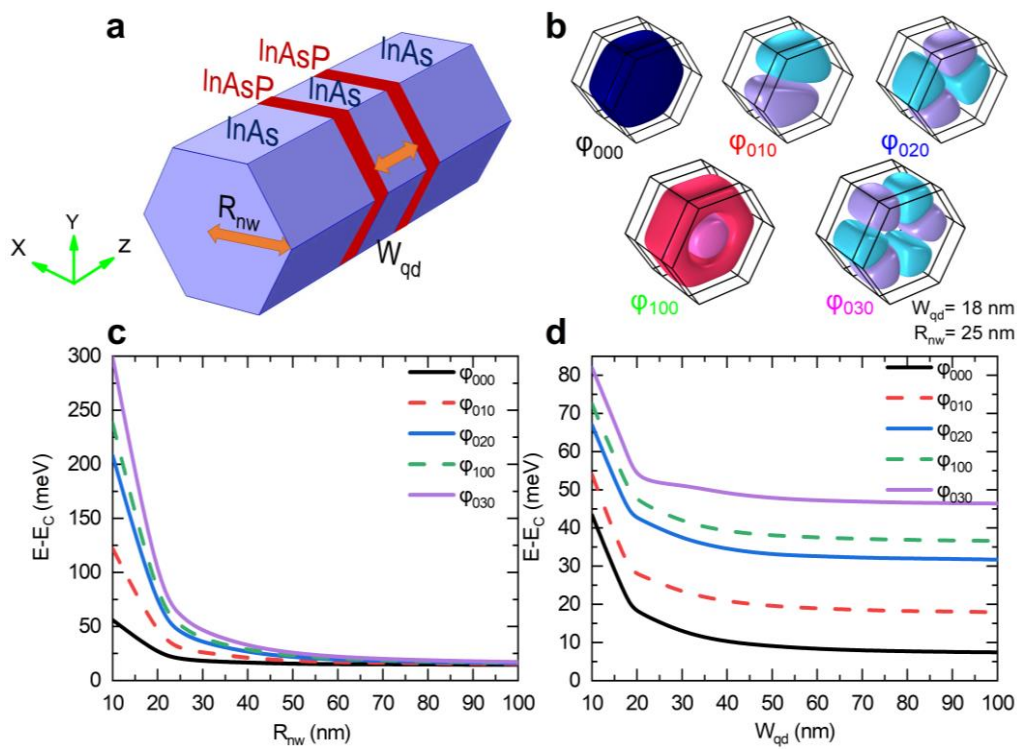

**Figure 2 a)** Three dimensional hexagonal wurtzite structure of the axially grown QD-NW, employed in our numerical simulations. **b)** Distribution of the electron wave function (orbital configuration) for the first five energy states of a QD, having the following dimensions: NW radius ($R_{nw}$= 25 nm) and QD width along the growth direction ($W_{qd}$ = 18 nm). **c),d)** Energy of the electronic states localized in the quantum dot, plotted as a function of $R_{nw}$ and $W_{qd}$ of the QD, respectively. Color lines (marked with letters, corresponding to panel b) present the evolution of the orbital energies. The ground state (black line) energy is $E_{000}$ > 55 meV when $R_{nw}$ approaches 10 nm, and it decreases noticeably while increasing the radius up to 50 nm ($E_{000}$ < 20meV). As expected, the confinement also depends on $W_{qd}$. The energies of the radially confined states (quantum number g=0, Fig. 2b) as a function of $W_{qd}$. Interestingly, while the energy spacing between g=0 levels steadily decreases with $R_{nw}$ (c), it remains almost unaltered when $W_{qd}$ increases from 40 nm to 100 nm (d), demonstrating the strong correlation of quantum confinement to the geometrical characteristics.

In order to couple the QD-NW based FET to the 0.6 THz field, the deeply sub-wavelength quantum dot element is asymmetrically integrated in a planar bow-tie antenna[33], with radius 210 µm (Fig. 1a). The asymmetry is ensured by the connection of the antenna arms to the S and lateral gates (G) electrodes.

We compute the eigenstates $|\varphi_{nlg}\rangle$ and eigenenergies $\varphi_{nlg}$ of axially grown QD by iteratively solving the coupled Schrödinger-Poisson equations; here the quantum numbers *n*, *l* and *g* distinguish between levels confined along the radial, angular, and growth directions, respectively. The QD model is geometrically sketched in Fig. 2a and the simulation results are shown in Figs 2b-2d. The computed energy spacing between the ground level and the first excited state, is ~ 9 meV for our specific QD geometry.

We investigate the electron transport through the SET by measuring the charge stability diagram at a heat sink temperature $T_0$ = 4.2 K. The detector in mounted in a pulse-tube optical cryostat with helium exchange gas (Janis Research) and the source-drain current ($I_{SD}$) is recorded as a function of both the source-drain bias ($V_{SD}$) and the gate voltage ($V_G$). The Coulomb blockade diagram (Figure 3a) shows diamonds corresponding to regions where the conductance is suppressed, since single-electron tunneling processes are not allowed and the electron occupation (N) in the quantum dot is fixed. By spanning $V_G$ beyond the pinch-off voltage ($V_G$ = 4.5V), if both the thermal energy ($k_BT$) and the source-drain biasing energy ($eV_{SD}$) are smaller than the charging energy of the QD ($k_BT \leq eV_{SD} \leq e^2 / C_\Sigma$), a set of conduction peaks are visible, due to the progressive alignment of electrochemical potential on the S and D sides with the QD energy levels. Between two consecutive peaks, resonant tunneling through the barriers is inhibited and the charge stability condition is satisfied. Importantly, the transport characteristics of the SET in the Coulomb blockade regime, being related to the single particle energy states and to their mutual interactions, allows describing the electronic features of the device. In particular, employing the constant interaction model[47], we evaluate the capacitive coupling between the QD and the gate electrode ($C_{gd}$), the charging energy, and the energy levels spectrum (Supporting information). Following these arguments, we extract from Fig. 3a: $e^2/C_\Sigma$ = 6.9 meV, an energy level spacing $\Delta E$ = 7.0 meV (in reasonable agreement with the value obtained by our Schrödinger-Poisson model), $C_\Sigma$ = 55 aF, $C_{gd}$ = 1.1 aF and $\alpha_G = C_{gd}/C_\Sigma$ = 18.8 mV/V.

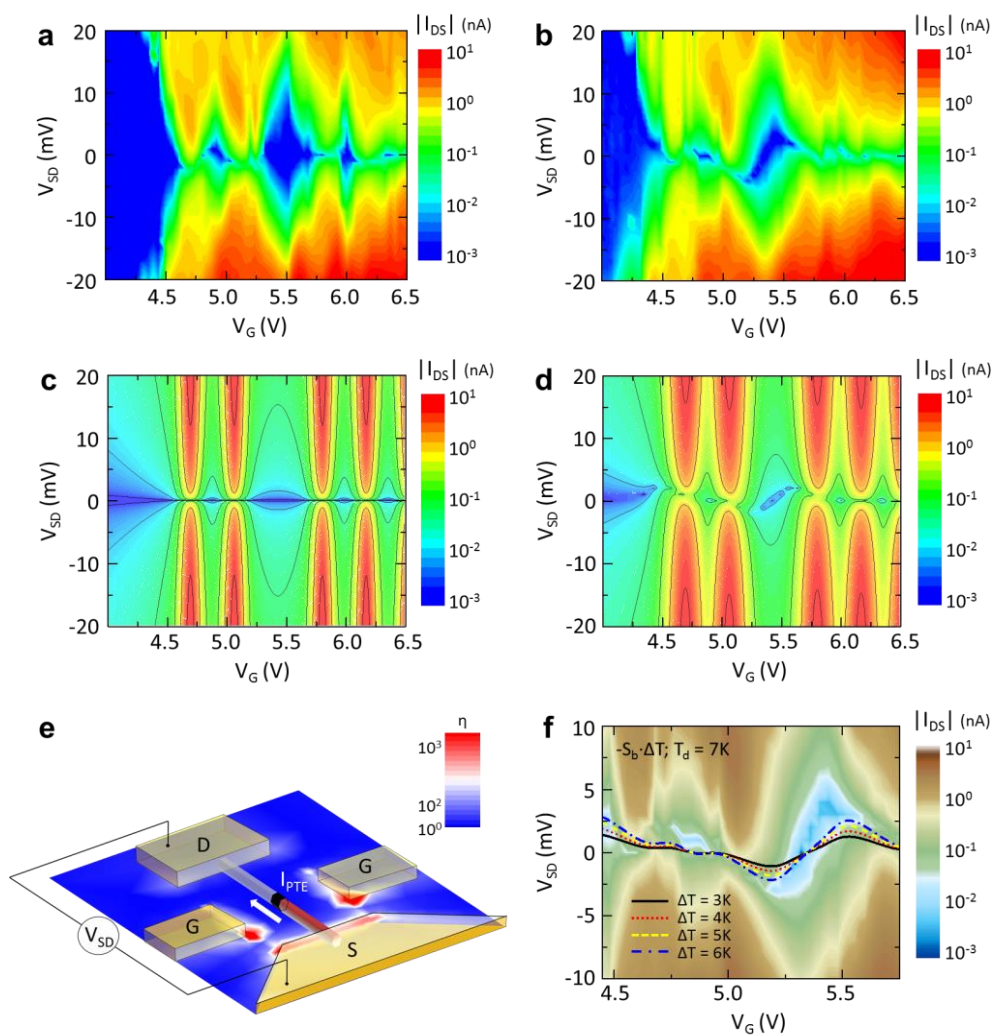

**Figure 3** Low temperature electrical transport through the single electron transistor in the dark (a,c) and illuminated (b,d) states. The absolute drain-source current $|I_{DS}|$ is plotted as a function of $V_{SD}$ (vertical axis) and $V_G$ (horizontal axis) in logarithmic scale. (a) Experimental dark state showing a typical Coulomb charge stability diagram, where the characteristic diamonds corresponding to a fixed electron population N=1,2,3 are resolved. (b) $|I_{DS}|$ map measured when the 0.6 THz radiation is illuminating the device, showing smeared and tilted Coulomb diamonds. (c,d) Theoretical charge stability diagrams. (c) Calculated THz-off map with the thermoelectric model, assuming the dot temperature $T_d$ equal to the heat-sink temperature $T_d=T_0 = 4$ K. (d) Calculated evolution of the Coulomb blockade diagram under illumination: a longitudinal source-drain thermal gradient $\Delta T = 5$ K and a global heating of the dot $T_d = 7$ K are assumed. (e) Schematic diagram of the detection mechanism: the sign of the photothermoelectric (PTE) contribution to the current is determined by the direction of the thermal gradient and by the sign of the Seebeck coefficient of the QD. The background color map is a numerical simulation of the electromagnetic energy density enhancement ($\eta$) under THz illumination. (f) Overlay between the experimental $|I_{DS}|$ color map and the PTE electromotive force $V_{PTE} = -S_b\Delta T$, calculated assuming different THz-induced $\Delta T$= 3 K, 4 K, 5 K, 6 K and for $T_d$=7 K.

Once funneling the THz beam onto the detection element (see Supporting Information), the retrieved *dc* current map appears visibly altered (Fig. 3b). The clearest features are the distorted Coulomb diamonds and the shift of the current peaks towards larger $V_G$ as compared to the corresponding peaks measured without THz radiation (Fig. 3a). In particular, the current vanishes to zero at different $V_{SD}$ and for different $V_G$, very differently from what happens in the dark (Fig. 3a), where, for any given $V_G$, it approaches zero at $V_{SD}= 0V$.

We ascribe this behavior to the photothermoelectric effect, generated by the thermal gradient ($\Delta T$) between the left and right side of the QD-NW: the bow-tie antenna funnels the radiation asymmetrically, producing a field enhancement, that is mainly confined between the S and G electrodes, leaving the drain-side colder with respect to the source-side of the QD.

Figure 3e schematically displays the detection mechanism, overlaid to the numerically simulated (finite element method, Comsol Multiphysics) electromagnetic energy enhancement ($\eta$) induced by the THz beam at the source-side of the NW. In this picture, the total current flowing along the SET channel $I_{DS} = \sigma_0 (V_{SD} + S_b \Delta T)$ where $\sigma_0 = dI_{SD}/dV_{SD}$ is the electronic conductance, $S_b$ represents the QD Seebeck coefficient (or thermopower). Under illumination, the distortion of the Coulomb diamonds is given by the additional electromotive force $V_{PTE} = S_b \Delta T$, which is identified in the current map (Figure 3b) by the $V_{SD}$ values at which the current vanishes along the gate voltage sweep (it takes a finite $V_{SD}$ to counterbalance $V_{PTE}$). Therefore, the amplitude and sign of $V_{PTE}$ are ultimately determined by the QD thermopower $S_b$. It is worth noticing that, by taking advantage of the PTE mechanism, our scheme allows for zero-bias, zero-dark current operation, limiting the detector noise to intrinsic charge or temperature fluctuations.

Such a qualitative interpretation of the PTE detection is then supported by the quantitative comparison of the experimental data with the outcome of a numerical model of charge and heat transport through the QD, based on the Landauer approach[49,50], to evaluate its thermoelectric properties, starting from electrical parameters that mimic our experimental configuration (see Supporting Information). The results, shown in Figure 3c, exhibit a good agreement with the experimental data (Figure 3a). When the THz beam is illuminating the detector, we expect that a thermal gradient $\Delta T$ between the left (S) and right (D) leads of the QD is established, as a consequence of the asymmetric coupling provided by the bow-tie antenna (Figure 3e). In addition to $\Delta T$, the evaluation of the theoretical map for the illuminated case (Figure 3d) also considers a global heating of the dot ($T_d > T_0$, where $T_d$ is the dot temperature). This accounts for the fact that the QD, being approximately located at the center of the SET, will reach, at steady state, an intermediate temperature between those of the S and D extremes. By comparing the model with the experimental results, we estimate an overall increase of $T_d$ from 4 K ($T_0$) in the dark state to 7 K in the illuminated case, with a THz-induced temperature difference $\Delta T = 5$ K between the leads. The increase of $T_d$ has the effect of

This is the authors' version of the article submitted to *Nano Letters* and accepted for publication
https://doi.org/10.1021/acs.nanolett.1c02022
increasing the coulomb peaks broadenings (see Supporting Information), whereas ΔT gives rise to a clear Seebeck effect, which corresponds to the additional electromotive force ($-S_b\Delta T$) acting on the dot. Such a picture is nicely captured by the superposition of the $|I_{DS}|$ map in the illuminated case (THz on) with the theoretical contour lines corresponding to $I_{DS} = 0$ ($V_{SD} = -S_b\Delta T$), calculated for different values of ΔT (Fig. 3f).

We ascribe the small discrepancy between the proposed model and the experimental results to the possible contribution of the photogating effect[51] and to the gate voltage and temperature dependent conductivity of the InAs NW segments on the left and right side of the dot, which is not taken into account. The developed simplified thermoelectric model however clearly highlights the main thermal effects governing the photodetection process.

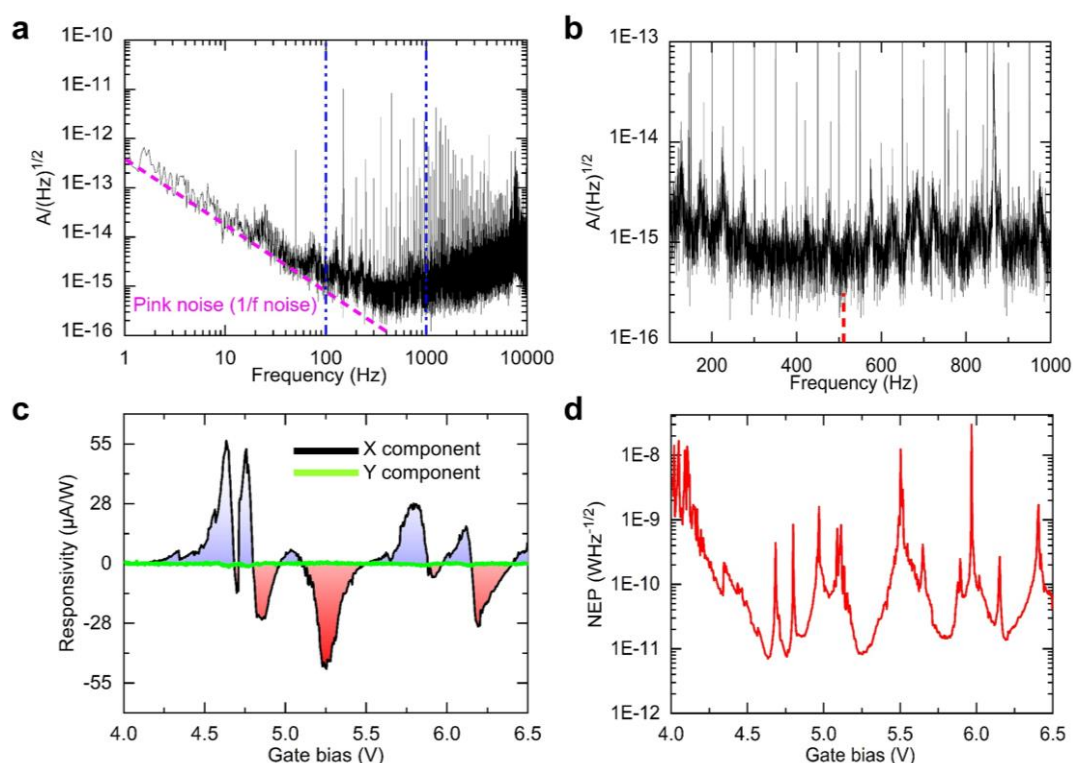

**Figure 4 a)** Noise Spectral Density (NSD) as a function of frequency measured by a spectrum analyzer. Pink dashed line is referring to the predicted 1/f noise contributor. The dotted blue lines mark the region of low noise. **b)** zoom of the NSD of panel a) plotted in the 100 Hz-1 kHz range while the red dashed line refers to the frequency equal to 518 Hz; **c)** Responsivity measured under illumination at 0.6 THz, at low temperature (T = 4.2 K), plotted as a function of $V_G$, while keeping $V_{SD}$ fixed at zero and measuring the X and Y components of the lock in amplifier. **d)** Noise equivalent power (NEP) plotted as a function of $V_G$.

We finally extrapolate the gate bias dependent detector responsivity ($R_a$) in order to evaluate the main figures of merit of the devised QD NW photodetector (see Supporting Information). As shown in Figure 4b, the observed dependence of $R_a$ from $V_G$ is correlated with the sharp

This is the authors' version of the article submitted to *Nano Letters* and accepted for publication
https://doi.org/10.1021/acs.nanolett.1c02022
transport peaks retrieved via transport experiments (Fig. 3b). In particular, the multiple sign changes of the photoresponse reflect the sign changes of the thermopower $S_b$ around each Coulomb peak and $R_a$ reaches a maximum value of 55 µA/W. The corresponding noise equivalent power (NEP)[52] is determined by the ratio between the noise spectral density (NSD) and $R_a$. The NSD is measured by connecting D to a Dacron dynamic spectrum analyzer (model Photon) (Figure 4a), while keeping S grounded and $V_G$=5V. At low frequency ($f$ < 200 Hz) the noise spectrum is mostly following $1/f^b$ trend (0<b<2 pink noise), whereas at higher frequencies ($f$ > 200 Hz) it is dominated by instrumental noise, slightly increasing as a function of frequency. Significantly, a local minimum in the noise figure occurs at 518 Hz, *i.e.* the frequency of the employed mechanical chopper. The contribution of the thermal (Johnson-Nyquist) noise in our system is ~0.5 × $10^{-15}$ A/$Hz^{1/2}$ for $V_G$ = 5V and $V_{SD}$ = 0V. The bias dependent NEP (Fig. 4c) shows a sequence of minima corresponding to the peaks observed in the responsivity curve. A minimum NEP of 8 pW$Hz^{-1/2}$ is reached (Fig. 4d). Although it appears still distant from the NEPs of commercially available cooled detectors as hot electron bolometers (NEPs~$10^{-19}$-$10^{-20}$ W$Hz^{-1/2}$),[15] the envisioned optimization guidelines (see Supporting information), combined with its zero-dark-current and inherent quantum nature make the proposed technology extremely appealing for quantum technology oriented applications.

We then estimate the temporal response of the proposed QD-NW PTE device through the analysis of its transport characteristics. As demonstrated experimentally, the timescales governing the heating/cooling dynamics of the carrier density in InAs NWs falls in the 40 fs - 4 ps range.[53] These timescales are much faster than the detector rise/fall times, which are limited by its electrical time constant $\tau_{RC} = R_tC_t$ ~ 1ns - 10 ns, where $R_t$ (~1MΩ - 10MΩ) and $C_t$ (~1 fF, includes the bow-tie shunt capacitance, simulated using COMSOL Multiphysics, electrostatic module) are the photodetector resistance and capacitance, respectively.

The achieved performances, combined with the extreme versatility of the QD-NW platform in terms of geometry and chemical composition, and with the intrinsically broadband and zero-bias nature of the PTE detection mechanism, unveiled through the choice of the impinging frequency - NW QD geometry combination, leave room for substantial improvements of the proposed quantum detection concept. For example, we envision that an optimization strategy for the PTE conversion shall proceed towards the reduction of the tunnel coupling between the dot and the leads[31], the engineering of the energy levels spacing and the exploitation of quantum phenomena, e.g. Kondo effect[54], in search of a balance between

1111

This is the authors' version of the article submitted to *Nano Letters* and accepted for publication
https://doi.org/10.1021/acs.nanolett.1c02022

detector sensitivity and speed (see Supporting Information file). Therefore, the reported results open a solid perspective for the combination of THz technology and few-electron physics to address some of the major challenges of quantum science, as quantum key distributions, quantum communications, and quantum sensing, where sub-shot noise NEPs combined with large quantum efficiencies are required. Furthermore, our work provides a clear understanding of the broadband PTE driven photoresponse in a QD-NW architecture, offering a framework for disentangling the different physical phenomena that would occur when the impinging photon energy matches the QD level spacing. However, the flexibility offered by quantum engineering to optimize device trasport and optical properties while simultaneously matching the photon energy with the QD energy level spacing, makes our QD InAs/InAsP heterostructured-nanowires an ideal building block in quantum-optics and nanophotonic applications, requiring a precise control of individual photon paths.

**Supporting Information**

Nanofabrication, Electromagnetic simulation of the antenna, Numerical model of transport, Evaluation of the photothermoelectric response, optical experiment, quantum engineering roadmap.

**Acknowledgements**

This work is supported by the European Research Council through the ERC Consolidator Grant (681379) SPRINT, by the European Union through the Marie Curie H2020-MSCA-ITN-2017, TeraApps (765426), by the SUPERTOP project, QUANTERA ERA-NET Cofound in Quantum Technologies grant (731473), and by the FET-OPEN project AndQC (828948). M.A., L.V. and M.S.V. acknowledge experimental support from Francesco Rossella. V.Z. and L.S. acknowledge Francesca Rossi for the TEM and EDX analysis. G.M. thanks Giuseppe Bevilacqua, Alessandro Cresti, and Giuseppe Grosso for useful discussions. M.A. and D.C. acknowledge experimental support from Cedric Bray. The work is also partially supported by the "International Research Agendas" program of the Foundation for Polish Science (CENTERA No. MAB/2018/9).